\documentclass[twocolumn]{aastex6}
\begin{document}

\title{ALMA Observations of the Solar Chromosphere on the Polar Limb}

\author{Takaaki Yokoyama\altaffilmark{1}}
\author{Masumi Shimojo\altaffilmark{2,3}}
\author{Takenori J. Okamoto\altaffilmark{2}}
\author{Haruhisa Iijima\altaffilmark{4}}
\altaffiltext{1}{Department of Earth and Planetary Science, The University of Tokyo, 
7-3-1 Hongo, Bunkyo-ku, Tokyo 113-0033, Japan}
\altaffiltext{2}{National Astronomical Observatory of Japan, NINS, Mitaka, Tokyo, 
181-8588, Japan}
\altaffiltext{3}{Department of Astronomical Science, SOKENDAI 
(The Graduate University of Advanced Studies), Mitaka, Tokyo, 181-8588, Japan}
\altaffiltext{4}{Institute for Space-Earth Environmental Research, Nagoya University, 
Furocho, Chikusa-ku, Nagoya, Aichi 464-0814 Japan}

\begin{abstract}
We report the results of the Atacama Large Millimeter/sub-millimeter 
Array (ALMA) observations of the solar chromosphere on the southern polar limb.
Coordinated observations with the Interface Region Imaging Spectrograph 
(IRIS) are also conducted. 
ALMA provided unprecedented high spatial resolution
in the millimeter band ($\approx$ 2.0\arcsec) 
at 100 GHz frequency with a moderate
cadence (20 s). The results are as follows: 
(1) The ALMA 100 GHz images show saw-tooth patterns on the limb, and
a comparison with SDO/AIA 171\AA\ images shows
a good correspondence of the limbs with each other.
(2) The ALMA 100 GHz movie shows a dynamic thorn-like structure
elongating from the saw-tooth patterns on the limb,
with lengths reaching at least 8\arcsec, thus suggesting jet-like activity
in the ALMA microwave range.
These ALMA jets are in good correspondence with IRIS jet clusters.
(3) A blob ejection event is observed. By comparing with the IRIS 
Mg II slit-jaw images, the trajectory of the blob is located along the spicular 
patterns. 

%The ejection is accompanied by a brightening jet event at the footpoint area.
\end{abstract}

\keywords{The Sun --- chromosphere --- magnetohydrodynamics}

\section{Introduction} \label{sec:intro}

The solar chromosphere is a region where many types of 
magnetohydrodynamic (MHD) phenomena can be observed 
such as brightening, eruptions, jets, and waves.
It has been mainly observed at visible or ultraviolet
wavelengths corresponding to absorption or emission lines.
It can also be observed in
microwave bands that correspond to emission mainly of the
thermal continuum from plasma.
%in the local thermodynamic equilibrium (LTE).
We can diagnose the electron density or temperature of the plasma.
According to the solar atmospheric
models by \citet{vernazza1981ApJS...45..635V}, the altitude 
of the 100 GHz emission is between 1300 and 2000 km
(see also \citealt{avrett2008ApJS..175..229A}).
During flares, 
non-thermal synchrotron emission 
in the optically thin range could appear and may provide information
of accelerated particles and magnetic fields.
The formation heights of the flare free-free ALMA continuum have been
computed by \cite{heinzel2012SoPh..277...31H}.

The Nobeyama Radioheliograph (NoRH, \citealt{nakajima1994IEEEP..82..705N})
is a microwave instrument dedicated for solar
observations at 17 and 34 GHz wavebands, which captures full disk
images of the Sun. Flares (\citealt{yokoyama2002ApJ...576L..87Y},
\citealt{asai2013ApJ...763...87A},
\citealt{minoshima2009ApJ...697..843M}), 
plasma eruptions 
(\citealt{gopalswamy2003ApJ...586..562G})
and long-term solar activity (\citealt{shimojo2013PASJ...65S..16S})
have been studied with this instrument. However, the spatial resolution
of NoRH is only 10\arcsec\ and 5\arcsec\ for 17 and 34 GHz, respectively,
which is insufficient
to investigate fine structures in the chromosphere.

The Atacama Large Millimeter/sub-millimeter
Array (ALMA, \citealt{wootten2009IEEEP..97.1463W}) 
has opened a new solar observational window with its excellent capabilities
that achieve a spatial resolution up to 1 arcsecond and is expected
to eventually be upgraded to sub-arcsecond resolution.
Some studies have already been published using ALMA science verification data
(\citealt{shimojo2017ApJ...848...62S},
\citealt{iwai2017ApJ...841L..20I},
\citealt{bastian2017ApJ...845L..19B},
\citealt{loukitcheva2017ApJ...850...35L}
).
From ALMA Cycle-4 operations, solar observations became accessible
on a proposal basis. Using this opportunity, we observed a 
polar region of the Sun with ALMA. 

One of interesting subjects on  the chromospheric dynamics
is the generation of jets,
such as spicules in quiet regions and coronal holes
and dynamic fibrils in active regions 
(\citealt{okamoto2011ApJ...736L..24O} and references therein).
Plasma flows with super-sonic speed are believed to be ejected via the
interaction of shock waves with the transition region contact surface
(\citealt{suematsu1982SoPh...75...99S,hollweg1982ApJ...257..345H}). 
These shock waves are generated by linear or non-linear mode conversion
through wave propagation after the generation at the photospheric level.
The original forms at the surface are magneto-sonic or Alfvenic waves,
which are under the strong influence of magnetic fields.
Moreover, during the propagation, the shocks suffer from the
radiative cooling of a plasma in non local termodynamic equiillibrium (non-LTE).
Thus, the jet production, propagation, and energy-loss
provide a clue for 
understanding the transport of energy to the corona.

In this paper, we report on our observations of the solar-limb chromosphere
using ALMA data. 
By comparing with ultraviolet data obtained by 
the Interface Region Imaging Spectrograph (IRIS), we found a correspondence
in the dynamic features in the ALMA data with IRIS spicules.
Among such jet activities, one event was observed that had 
a blob-like ejection.
This is the first ALMA report on the jet activities in the chromosphere.
We report
the observations in Section \ref{sec:observations}, 
results in Section \ref{sec:results},
discussion in Section \ref{sec:discussion},
and conclusion in Section \ref{sec:conclusion}.

\section{Observations} \label{sec:observations}

The observations were performed in $\approx$ 1 h around 14:30 UT on 
2017 April 29.
The target region was the solar south pole. ALMA obtained Band-3 data
in this time period within the 100 GHz frequency range. 
The instrumental integration time was 2 s.
Owing to weather conditions,
a Band-6 observation that was also proposed at 239 GHz was cancelled. 
Because all of the data in the spectrum window is used for image synthesis, 
the observing frequency of the image (100 GHz) is the same as that 
of the local oscillator. The major and minor axes of the synthesized beam 
are 2.56\arcsec\ (1.86 Mm on the Sun) 
in the east-west direction and 1.60\arcsec\ (1.16 Mm)
in the north-south direction, respectively.

The ALMA data is calibrated with the nominal processes 
developed for solar observations
in \citet{shimojo2017SoPh..292...87S}. 
In addition to those calibrations, to increase the image quality
(in particular, the contrast), we synthesize each image
with an integration time of 20 s after
performing a following two-step self-calibration 
for the antenna's phases.
In the first step, 
we construct a calibration model using an image synthesized by 
data from the entire observation period. This reduces the jitter 
motion caused by variation in sky seeing. 
In the second step, we repeated the same procedure but
synthesized images every 20 s and used each of them 
with a 20-s cadence as a calibration model for each data slot. 
Despite these calibrations,
it is still difficult to determine the absolute brightness temperature
in the ALMA data 
(For detailed discussion on limb artifacts and
the speicial treatment, see  \citealt{shimojo2017SoPh..292...87S}).
Therefore, we assume that the brightness of the
solar disk is 7000 K (\citealt{avrett2008ApJS..175..229A}).
Note that this assumption does not significantly
affect our results because the interpretation is limited
to the relative enhancements beyond the background signal.

A coordinated observation with 
%the Interface Region Imaging Spectrograph 
IRIS (\citealt{depontieu2014SoPh..289.2733D})
was executed so that ultraviolet data in the same field of view was available.
IRIS obtained slit-jaw images (SJIs) of Si IV and Mg II lines
with a cadence and a spatial resolution of 19~s and
$0.33$\arcsec\--$0.40$\arcsec\ (240\,--\,290~km on the Sun), respectively.
The Atmospheric Imaging Assembly (AIA; 
\citealt{lemen2012SoPh..275...17L}) on board 
the Solar Dynamics Observatory (SDO) 
provided context filtergrams of multiple wavelengths with a cadence of 12 s.

\section{Results} \label{sec:results}

\begin{figure*}[ht!]
\plotone{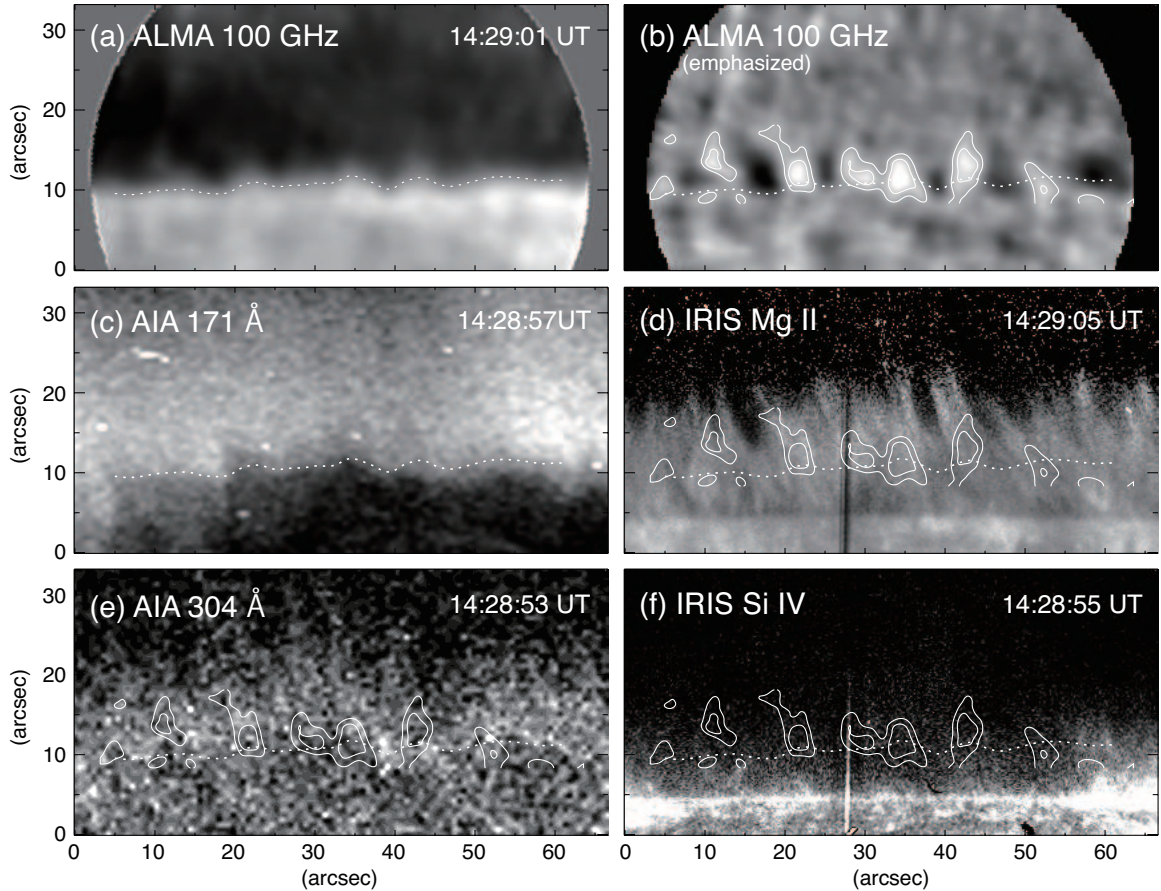}
\caption{Observation results. Solar south is up. 
(a) Brightness temperature $T_\mathrm{b}$ and (b) relative brightness 
$\tilde{T}_\mathrm{b}$
(see the text for definition) of ALMA Band-3 at 100 GHz,
(c) SDO/AIA 171\AA\ band images, (d) IRIS/SJI Mg II, (e) SDO/AIA 304\AA\, and
(f) IRIS/SJI Si IV.
Solid contours represent $\tilde{T}_\mathrm{b}=$ 1.02 and 1.04.
The dotted contour represents $T_\mathrm{b}=$ 6500 K.
(movie available)
\label{fig:fullfov}}
\end{figure*}

Figure \ref{fig:fullfov} shows the observational results. 
The ALMA 100 GHz image (panel a) shows a saw-tooth pattern on the limb. 
There is a good correspondence in the limb location between 
the ALMA data and the AIA 171\AA\ images (see the dotted line in panel c). 
This means that the AIA 171\AA\ limb is a manifestation of the absorption by
the chromospheric materials observed in the ALMA data. 

Panel b shows the distribution of the relative brightness in the 
ALMA 100 GHz data defined as follows:
\begin{equation}
\tilde{T}_\mathrm{b}=T_\mathrm{b}/\langle T_\mathrm{b} \rangle
\end{equation}
where
\begin{equation}
\langle {T}_\mathrm{b}\rangle =\frac{1}{\Delta t}\int T_b dt
\end{equation}
is a temporally averaged brightness over the entire observed period 
($\Delta t \approx 1 \ \mathrm{h}$).
The ALMA 100 GHz relative brightness and 
its corresponding movie (panel b) show a dynamic 
thorn-like structure
coming from the saw-tooth patterns on the limb. Because these
structures show dynamic motion with elongation and
shortening, hereafter we call them ALMA jets.
Their length reaches at least 8\arcsec, suggesting jet-like activity
in the ALMA microwave range.
Unfortunately owing to the insufficient signal 
to noise ratio in the data beyond the limb particularly 
in the area more than 10\arcsec\ beyond the limb 
(see Section \ref{sec:discussion}), we can not recognize 
the maximum height of the elongated jets. The speed
of the elongation is 30--40 $\mathrm{km}\ \mathrm{s}^{-1}$.
The width of each jet is less than the spatial resolution of
the observation, i.e. less than $\approx$ 2\arcsec.

The IRIS/SJI Mg II images (panel d) clearly show the appearance of spicules on the
limb. Some of the spicular jets reach more than 20\arcsec\ beyond the Mg II limb.
It is noteworthy that the spicular jets show clustered structures, i.e.
multiple needle-like features display a collective up-and-down motion.
Each needle-like structure has a width less than 1\arcsec\ and each
cluster has a few to several needle-like structures.  In comparison with the
ALMA data (contours in panel d), the ALMA jets are
in good agreement with the IRIS jet clusters.

The IRIS Mg II limb (panel d) is $\approx$ 5\arcsec\ below the ALMA limb.
Again, this confirms the distribution of cool chromospheric plasmas between
the IRIS Mg II and ALMA limbs (For IRIS limb analysis,
see \citealt{alissandrakis2018SoPh..293...20A}).
In the IRIS/SJI Si data (panel f), one can observe the activities on the
solar surface inside the limb. Some of them correspond to the footpoints
of the Mg II spicules. The AIA 304\AA\ data (panel e) has 
a low signal-to-noise
ratio and is difficult to extract information on jet activities.

\begin{figure*}[ht!]
\plotone{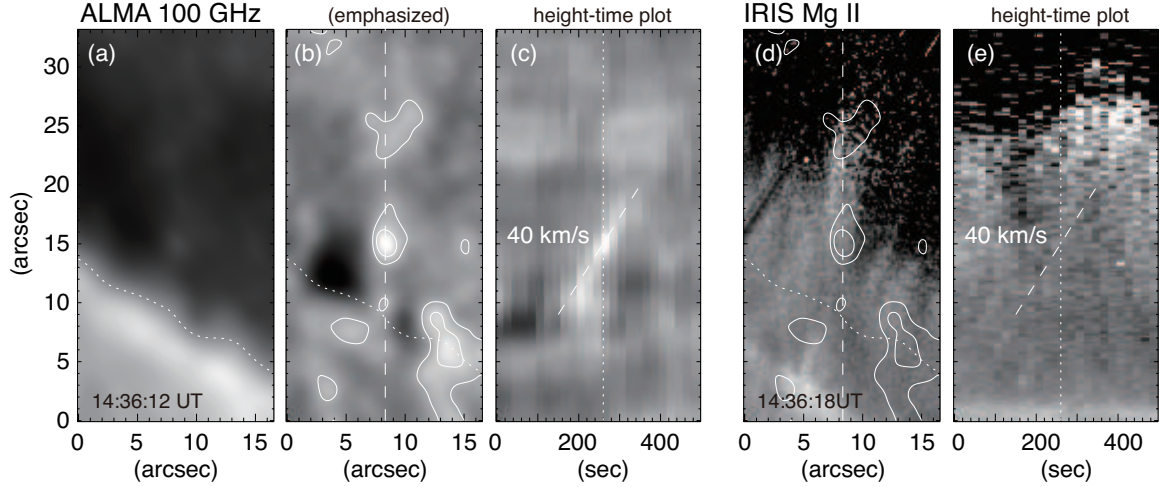}
\caption{
An ALMA blob ejection event associated with a tall IRIS Mg II jet.
(a) Brightness temperature $T_\mathrm{b}$ and 
(b) relative brightness $\tilde{T}_\mathrm{b}$
of ALMA 100 GHz at the moment of the ALMA blob ejection.
(c)  Distance along the trajectory versus time stack  plot 
of ALMA relative brightness along the dashed line in panel (b).
The horizontal (vertical) 
axis is for time (space along the jet). (d) and (e) for
IRIS Mg II intensity as (a) and (c).
Solid contours in (b) and (d) are for $\tilde{T}_\mathrm{b}=$ 1.02 and 1.04.
Dashed lines in (c) and (e) indicate the slope for a speed of 40 km/s.
\label{fig:talljet}}
\end{figure*}

\begin{figure}[ht!]
\plotone{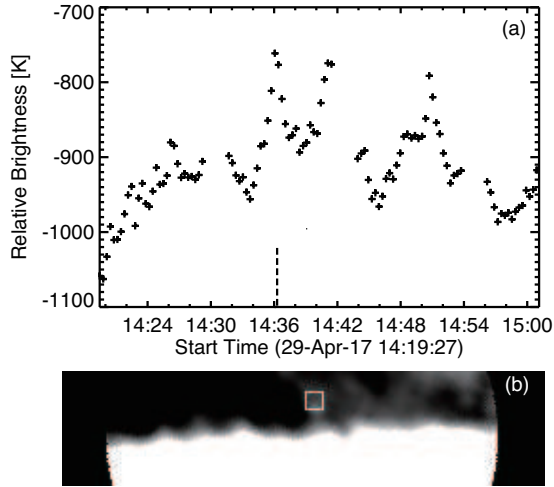}
\caption{
(a) Time profile of ALMA 100 GHz brightness temperature 
averaged over the area in the solid-line box
of (b). The vertical dashed line indicates the time of the
blob ejection event. Note that the average over an image at each time
is set to be zero in this plot.
\label{fig:time_prof}}
\end{figure}

In one of the tall jets in IRIS Mg II, an ALMA blob ejection was found. 
From Figure \ref{fig:talljet}, one can find the trajectory of the blob
co-located with the Mg II jet.  The brightness enhancement in ALMA
images is by 135 K beyond the background
(Figure \ref{fig:time_prof}).
The blob shows a systematic proper motion at a speed of 40 km/s (panel c), 
which is similar to the IRIS jet speed (panel e).
There are no significant features in the IRIS Si or the AIA data sets
co-located with this blob event.
However, by the careful inspection of the IRIS Si data, we found
short ($\approx$ a few arcseconds) elongated structures at 
the base of this event. They show shifting from east to west changing 
its inclination angle. 

\section{Discussion} \label{sec:discussion}

\begin{figure*}[ht!]
\plotone{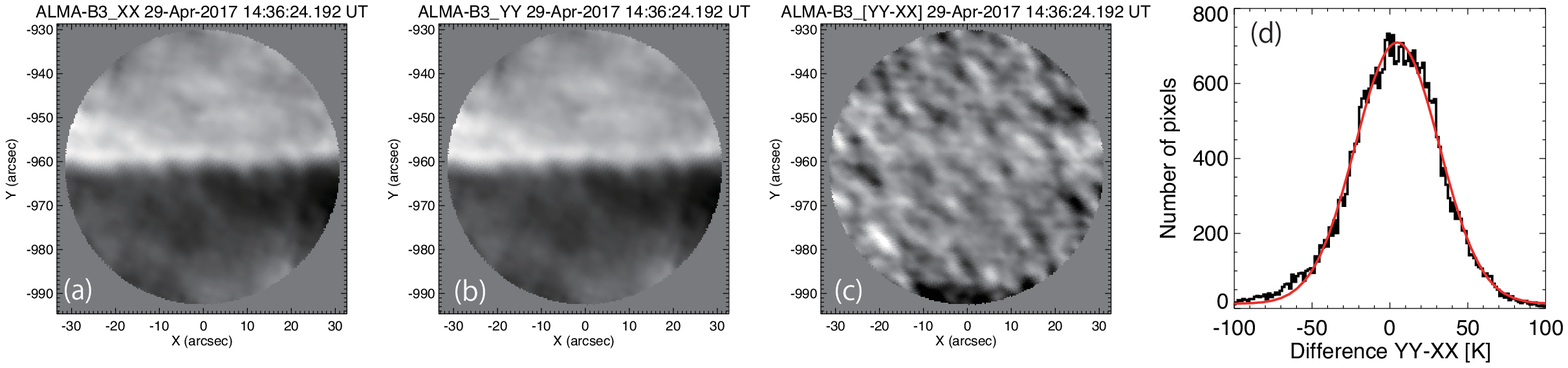}
\caption{
(a), (b) Synthesized snapshots from the two sets of correlations, XX and YY, 
respectively, using the two orthogonal linear polarization data. 
Solar south is down.
(c) is subtraction image and (d) is its histogram of XX from YY. 
\label{fig:noise}}
\end{figure*}

The noise level in the 100 GHz images is evaluated
using  the XX and YY cross correlations
(out of the four available, XX, YY, XY, and YX) of 
orthogonal linear polarizations (X and Y) measured by ALMA receivers.
In the absence of any flare emission, which is the case of our study,
solar mm/sub-mm emission is thermal with very weak net linear
polarization. Assuming the subtraction of brightness temperature 
obtained by each of the XX- and YY- correlations is zero, the remaining
signals can be regarded as noises (\citealt{shimojo2017SoPh..292...87S}). 
The measured level of the noise is 13.2 K (Figure \ref{fig:noise}). 
Although the lower part of the images (Figure \ref{fig:noise}a, and b) 
indicates the sky, we can observe brightness enhancement. 
The enhancement is caused by the artifacts described in 
\citet{shimojo2017SoPh..292...87S}.
Hence, we use only the data near the limb (image center) for the analysis,
where these artifacts have a weak influence on our results.

In terms of the blob ejection event, an estimation of the physical
variables may help its interpretation. Here we try to derive
the electron density based on an assumption of the emission via the thermal
free-free interaction, i.e. its optical thickness given as follows
(e.g. \citealt{dulk1985ARA&A..23..169D,lang1980afcp.book.....L}):
\begin{equation}
\begin{array}{lll}
\tau 
&=&
\frac{\mathrm{EM}}{(2\pi\nu)^2}\left[
\frac{32 \pi^2 e^6}{3 (2\pi)^{1/2} m^3 c}
\right]
\left(
\frac{m}{k_B T}
\right)^{3/2}
\frac{\pi}{3^{1/2}}\bar{g_{\mathrm{ff}}}
\\
&=&
1.77 \times 10^{-2}\
\mathrm{EM}\ T^{-3/2}\nu^{-2}
\bar{g_{\mathrm{ff}}}
%\\
%&=&
%2.98 \times 10^{-5}
%\left(\frac{\mathrm{EM}}{10^{27}\ \mathrm{cm}^{-5}}\right)
%\left(\frac{T}{10^{5}\ \mathrm{K}}\right)^{-3/2}
%\left(\frac{\nu}{100\ \mathrm{GHz}}\right)^{-2}
%\bar{g_{\mathrm{ff}}}
\end{array}
\end{equation}
where
\begin{equation}
\begin{array}{lll}
\bar{g_{\mathrm{ff}}}
&=&
\frac{{3}^{1/2}}{\pi}
\ln\frac{(2 k_B T)^{3/2}}{\pi e^2 \nu m^{1/2} \gamma^{5/2}}
\\
&=&0.551\left[
17.7+\ln\left(\frac{T^{3/2}}{\nu}\right)
\right]
%\\
%&=&0.551\left[
%\frac{3}{2}\ln\left(\frac{T}{10^5\ \mathrm{K}}\right)
%-\ln\left(\frac{\nu}{100\ \mathrm{GHz}}\right)
%+9.66
%\right]
\end{array}
\end{equation}
is a velocity averaged Gaunt factor, 
$\mathrm{EM}\approx n^2 L$ emission measure. $\nu$ is the observing frequency,
$T$, $n$, $L$ are the temperature,
%in unit of Kelvin, 
electron density, and
the distributing length along the line of sight of the emitting plasmas,
respectively,
$m$, $k_B$, $c$, and $e$ are the electron mass, the Boltzmann constant,
the speed of light, and the elementary charge, respectively, and
$\gamma=1.781$ is a constant.
Considering the observed brightness temperature $T_b=135\ \mathrm{K}$ and,
assuming a small optical thickness, 100\% filling factor, and 
$L \approx 1.5\arcsec$ approximated with the width of the
observed blob, and $\tau \approx T_b/T$, we obtain 
$n=(4.6-8.4) \times 10^{9}\ \mathrm{cm}^{-3}$ in the range of
$T=(10^4- 4\times 10^5)\ \mathrm{K}$.
This number is a factor of one order
of magnitude smaller than those for spicules obtained through the analysis of
the IRIS spectrum data by \citet{alissandrakis2018SoPh..293...20A}
(see also \citealt{beckers1972ARA&A..10...73B};
\citealt{krall1976SoPh...46...93K}). The reason for this discrepancy might
result from our assumed filling factor of 100\%.
Because our measurement of the blob size as $L \approx 1.5\arcsec$
is comparable with the spatial resolution of the observation, 
it is highly possible that $L$ is overestimated so that affecting
the filling volume estimation in every dimension.
To make this result improved, 
we need at least higher resolution observations by ALMA.

For a production mechanism of tall spicules, \citet{iijima2017ApJ...848...38I}
proposed a model in which jets are driven by the shock waves originated
from the highly non-linear large amplitude Alfven waves in the chromosphere
(\citealt{iijima2015ApJ...812L..30I}, see also 
\citealt{kudoh1999ApJ...514..493K}). 
The other proposed energy sources are
the acoustic waves generated by the photospheric convection motion
(\citealt{thomas1948ApJ...108..130T},
\citealt{osterbrock1961ApJ...134..347O},
\citealt{uchida1961PASJ...13..321U}) and
acoustic waves generated by magnetic reconnection 
(\citealt{pikelner1971CoASP...3...33P},
\citealt{uchida1969PASJ...21..128U}).
The magnetic reconnection model is advantageous because the
chromospheric jets are accelerated by both the shock wave
and the Lorentz force (\citealt{takasao2013PASJ...65...62T}).
Using the available data in the present observations, we could not confirm
the proposed model. For further
investigation of the jet generation mechanisms, high resolution images
around the base of jets are required. This is achievable using 
other high frequency modes of ALMA, such as Band-6 (239 GHz) or beyond in the 
subsequent observational cycles.

\section{Conclusion} \label{sec:conclusion}

We report the results of the ALMA observations of the solar 
chromosphere on the southern polar limb. The results are as follows: 
(1) The ALMA 100 GHz image shows saw-tooth patterns on the limb. 
The comparison with AIA 171\AA\ images shows
a good correspondence of the limbs in both data sets.
(2) The ALMA 100 GHz movie shows a dynamic thorn-like structure
rising from the saw-tooth patterns on the limb. 
Their length reaches at least 8\arcsec, thus suggesting jet-like activities
in the ALMA microwave range.
These ALMA jets are in good agreement with IRIS jet clusters.
(3) A blob-ejection event is found. By comparing with the IRIS 
Mg II SJIs, the trajectory of the blob is located along the spicular 
patterns. The signal level is 135 K at ten-sigma level.
The blob shows a systematic proper 
motion at a speed of 40 km/s, which is similar to the jet speed.

\acknowledgments
This paper makes use of the following ALMA data: ADS/JAO.ALMA\#2016.1.00201.S. 
ALMA is a partnership of ESO (representing its member states), NSF (USA) 
and NINS (Japan), together with NRC (Canada), MOST and ASIAA (Taiwan), 
and KASI (Republic of Korea), in cooperation with the Republic of Chile. 
The Joint ALMA Observatory is operated by ESO, AUI/NRAO and NAOJ.
IRIS is a NASA small explorer mission developed and operated
by LMSAL with mission operations executed at NASA Ames Research
center and major contributions to downlink communications funded
by ESA and the Norwegian Space Centre.
SDO is part of NASA's Living With a Star Program.
The authors are supported by JSPS KAKENHI Grant:
T.Y. is by JP15H03640, M.S. by JP17K05397, 
T.J.O. by JP16K17663 (PI: T.J.O.) and JP25220703 (PI: S. Tsuneta),
and H.I. by JP15H05816..

\bibliography{draftref}

\begin{thebibliography}{}
\expandafter\ifx\csname natexlab\endcsname\relax\def\natexlab#1{#1}\fi

\bibitem[{{Alissandrakis} {et~al.}(2018){Alissandrakis}, {Vial}, {Koukras},
  {Buchlin}, \& {Chane-Yook}}]{alissandrakis2018SoPh..293...20A}
{Alissandrakis}, C.~E., {Vial}, J.-C., {Koukras}, A., {Buchlin}, E., \&
  {Chane-Yook}, M. 2018, \solphys, 293, 20

\bibitem[{{Asai} {et~al.}(2013){Asai}, {Kiyohara}, {Takasaki}, {Narukage},
  {Yokoyama}, {Masuda}, {Shimojo}, \& {Nakajima}}]{asai2013ApJ...763...87A}
{Asai}, A., {Kiyohara}, J., {Takasaki}, H., {et~al.} 2013, \apj, 763, 87

\bibitem[{{Avrett} \& {Loeser}(2008)}]{avrett2008ApJS..175..229A}
{Avrett}, E.~H., \& {Loeser}, R. 2008, \apjs, 175, 229

\bibitem[{{Bastian} {et~al.}(2017){Bastian}, {Chintzoglou}, {De Pontieu},
  {Shimojo}, {Schmit}, {Leenaarts}, \&
  {Loukitcheva}}]{bastian2017ApJ...845L..19B}
{Bastian}, T.~S., {Chintzoglou}, G., {De Pontieu}, B., {et~al.} 2017, \apjl,
  845, L19

\bibitem[{{Beckers}(1972)}]{beckers1972ARA&A..10...73B}
{Beckers}, J.~M. 1972, \araa, 10, 73

\bibitem[{{De Pontieu} {et~al.}(2014){De Pontieu}, {Title}, {Lemen}, {Kushner},
  {Akin}, {Allard}, {Berger}, {Boerner}, {Cheung}, {Chou}, {Drake}, {Duncan},
  {Freeland}, {Heyman}, {Hoffman}, {Hurlburt}, {Lindgren}, {Mathur}, {Rehse},
  {Sabolish}, {Seguin}, {Schrijver}, {Tarbell}, {W{\"u}lser}, {Wolfson},
  {Yanari}, {Mudge}, {Nguyen-Phuc}, {Timmons}, {van Bezooijen}, {Weingrod},
  {Brookner}, {Butcher}, {Dougherty}, {Eder}, {Knagenhjelm}, {Larsen},
  {Mansir}, {Phan}, {Boyle}, {Cheimets}, {DeLuca}, {Golub}, {Gates}, {Hertz},
  {McKillop}, {Park}, {Perry}, {Podgorski}, {Reeves}, {Saar}, {Testa}, {Tian},
  {Weber}, {Dunn}, {Eccles}, {Jaeggli}, {Kankelborg}, {Mashburn}, {Pust},
  {Springer}, {Carvalho}, {Kleint}, {Marmie}, {Mazmanian}, {Pereira}, {Sawyer},
  {Strong}, {Worden}, {Carlsson}, {Hansteen}, {Leenaarts}, {Wiesmann},
  {Aloise}, {Chu}, {Bush}, {Scherrer}, {Brekke}, {Martinez-Sykora}, {Lites},
  {McIntosh}, {Uitenbroek}, {Okamoto}, {Gummin}, {Auker}, {Jerram}, {Pool}, \&
  {Waltham}}]{depontieu2014SoPh..289.2733D}
{De Pontieu}, B., {Title}, A.~M., {Lemen}, J.~R., {et~al.} 2014, \solphys, 289,
  2733

\bibitem[{{Dulk}(1985)}]{dulk1985ARA&A..23..169D}
{Dulk}, G.~A. 1985, \araa, 23, 169

\bibitem[{{Gopalswamy} {et~al.}(2003){Gopalswamy}, {Shimojo}, {Lu}, {Yashiro},
  {Shibasaki}, \& {Howard}}]{gopalswamy2003ApJ...586..562G}
{Gopalswamy}, N., {Shimojo}, M., {Lu}, W., {et~al.} 2003, \apj, 586, 562

\bibitem[{{Heinzel} \& {Avrett}(2012)}]{heinzel2012SoPh..277...31H}
{Heinzel}, P., \& {Avrett}, E.~H. 2012, \solphys, 277, 31

\bibitem[{{Hollweg}(1982)}]{hollweg1982ApJ...257..345H}
{Hollweg}, J.~V. 1982, \apj, 257, 345

\bibitem[{{Iijima} \& {Yokoyama}(2015)}]{iijima2015ApJ...812L..30I}
{Iijima}, H., \& {Yokoyama}, T. 2015, \apjl, 812, L30

\bibitem[{{Iijima} \& {Yokoyama}(2017)}]{iijima2017ApJ...848...38I}
---. 2017, \apj, 848, 38

\bibitem[{{Iwai} {et~al.}(2017){Iwai}, {Loukitcheva}, {Shimojo}, {Solanki}, \&
  {White}}]{iwai2017ApJ...841L..20I}
{Iwai}, K., {Loukitcheva}, M., {Shimojo}, M., {Solanki}, S.~K., \& {White},
  S.~M. 2017, \apjl, 841, L20

\bibitem[{{Krall} {et~al.}(1976){Krall}, {Bessey}, \&
  {Beckers}}]{krall1976SoPh...46...93K}
{Krall}, K.~R., {Bessey}, R.~J., \& {Beckers}, J.~M. 1976, \solphys, 46, 93

\bibitem[{{Kudoh} \& {Shibata}(1999)}]{kudoh1999ApJ...514..493K}
{Kudoh}, T., \& {Shibata}, K. 1999, \apj, 514, 493

\bibitem[{{Lang}(1980)}]{lang1980afcp.book.....L}
{Lang}, K.~R. 1980, {Astrophysical Formulae. A Compendium for the Physicist and
  Astrophysicist.} (Springer-Verlag Berlin Heidelberg New York)

\bibitem[{{Lemen} {et~al.}(2012){Lemen}, {Title}, {Akin}, {Boerner}, {Chou},
  {Drake}, {Duncan}, {Edwards}, {Friedlaender}, {Heyman}, {Hurlburt}, {Katz},
  {Kushner}, {Levay}, {Lindgren}, {Mathur}, {McFeaters}, {Mitchell}, {Rehse},
  {Schrijver}, {Springer}, {Stern}, {Tarbell}, {Wuelser}, {Wolfson}, {Yanari},
  {Bookbinder}, {Cheimets}, {Caldwell}, {Deluca}, {Gates}, {Golub}, {Park},
  {Podgorski}, {Bush}, {Scherrer}, {Gummin}, {Smith}, {Auker}, {Jerram},
  {Pool}, {Soufli}, {Windt}, {Beardsley}, {Clapp}, {Lang}, \&
  {Waltham}}]{lemen2012SoPh..275...17L}
{Lemen}, J.~R., {Title}, A.~M., {Akin}, D.~J., {et~al.} 2012, \solphys, 275, 17

\bibitem[{{Loukitcheva} {et~al.}(2017){Loukitcheva}, {Iwai}, {Solanki},
  {White}, \& {Shimojo}}]{loukitcheva2017ApJ...850...35L}
{Loukitcheva}, M.~A., {Iwai}, K., {Solanki}, S.~K., {White}, S.~M., \&
  {Shimojo}, M. 2017, \apj, 850, 35

\bibitem[{{Minoshima} {et~al.}(2009){Minoshima}, {Imada}, {Morimoto}, {Kawate},
  {Koshiishi}, {Kubo}, {Inoue}, {Isobe}, {Masuda}, {Krucker}, \&
  {Yokoyama}}]{minoshima2009ApJ...697..843M}
{Minoshima}, T., {Imada}, S., {Morimoto}, T., {et~al.} 2009, \apj, 697, 843

\bibitem[{{Nakajima} {et~al.}(1994){Nakajima}, {Nishio}, {Enome}, {Shibasaki},
  {Takano}, {Hanaoka}, {Torii}, {Sekiguchi}, {Bushimata}, {Kawashima},
  {Shinohara}, {Irimajiri}, {Koshiishi}, {Kosugi}, {Shiomi}, {Sawa}, \&
  {Kai}}]{nakajima1994IEEEP..82..705N}
{Nakajima}, H., {Nishio}, M., {Enome}, S., {et~al.} 1994, IEEE Proceedings, 82,
  705

\bibitem[{{Okamoto} \& {De Pontieu}(2011)}]{okamoto2011ApJ...736L..24O}
{Okamoto}, T.~J., \& {De Pontieu}, B. 2011, \apjl, 736, L24

\bibitem[{{Osterbrock}(1961)}]{osterbrock1961ApJ...134..347O}
{Osterbrock}, D.~E. 1961, \apj, 134, 347

\bibitem[{{Pikel'ner}(1971)}]{pikelner1971CoASP...3...33P}
{Pikel'ner}, S.~B. 1971, Comments on Astrophysics and Space Physics, 3, 33

\bibitem[{{Shimojo}(2013)}]{shimojo2013PASJ...65S..16S}
{Shimojo}, M. 2013, \pasj, 65, S16

\bibitem[{{Shimojo} {et~al.}(2017{\natexlab{a}}){Shimojo}, {Iwai}, {Asai},
  {Nozawa}, {Minamidani}, \& {Saito}}]{shimojo2017ApJ...848...62S}
{Shimojo}, M., {Iwai}, K., {Asai}, A., {et~al.} 2017{\natexlab{a}}, \apj, 848,
  62

\bibitem[{{Shimojo} {et~al.}(2017{\natexlab{b}}){Shimojo}, {Bastian}, {Hales},
  {White}, {Iwai}, {Hills}, {Hirota}, {Phillips}, {Sawada}, {Yagoubov},
  {Siringo}, {Asayama}, {Sugimoto}, {Braj{\v s}a}, {Skoki{\'c}}, {B{\'a}rta},
  {Kim}, {de Gregorio-Monsalvo}, {Corder}, {Hudson}, {Wedemeyer}, {Gary}, {De
  Pontieu}, {Loukitcheva}, {Fleishman}, {Chen}, {Kobelski}, \&
  {Yan}}]{shimojo2017SoPh..292...87S}
{Shimojo}, M., {Bastian}, T.~S., {Hales}, A.~S., {et~al.} 2017{\natexlab{b}},
  \solphys, 292, 87

\bibitem[{{Suematsu} {et~al.}(1982){Suematsu}, {Shibata}, {Neshikawa}, \&
  {Kitai}}]{suematsu1982SoPh...75...99S}
{Suematsu}, Y., {Shibata}, K., {Neshikawa}, T., \& {Kitai}, R. 1982, \solphys,
  75, 99

\bibitem[{{Takasao} {et~al.}(2013){Takasao}, {Isobe}, \&
  {Shibata}}]{takasao2013PASJ...65...62T}
{Takasao}, S., {Isobe}, H., \& {Shibata}, K. 2013, \pasj, 65, 62

\bibitem[{{Thomas}(1948)}]{thomas1948ApJ...108..130T}
{Thomas}, R.~N. 1948, \apj, 108, 130

\bibitem[{{Uchida}(1961)}]{uchida1961PASJ...13..321U}
{Uchida}, Y. 1961, \pasj, 13, 321

\bibitem[{{Uchida}(1969)}]{uchida1969PASJ...21..128U}
---. 1969, \pasj, 21, 128

\bibitem[{{Vernazza} {et~al.}(1981){Vernazza}, {Avrett}, \&
  {Loeser}}]{vernazza1981ApJS...45..635V}
{Vernazza}, J.~E., {Avrett}, E.~H., \& {Loeser}, R. 1981, \apjs, 45, 635

\bibitem[{{Wootten} \& {Thompson}(2009)}]{wootten2009IEEEP..97.1463W}
{Wootten}, A., \& {Thompson}, A.~R. 2009, IEEE Proceedings, 97, 1463

\bibitem[{{Yokoyama} {et~al.}(2002){Yokoyama}, {Nakajima}, {Shibasaki},
  {Melnikov}, \& {Stepanov}}]{yokoyama2002ApJ...576L..87Y}
{Yokoyama}, T., {Nakajima}, H., {Shibasaki}, K., {Melnikov}, V.~F., \&
  {Stepanov}, A.~V. 2002, \apjl, 576, L87

\end{thebibliography}

\end{document}